\documentclass{ws-ijmpa}
\usepackage[super,compress]{cite}
\usepackage{graphicx}

\usepackage{hyperref}

\usepackage{epsfig}
\usepackage{epstopdf}
\usepackage{color}

\newcommand{\vslash}[1]{#1 \hspace{-0.5 em} /}

\usepackage{multirow}
\usepackage{booktabs}

\begin{document}
\markboth{I.V. Truten and A.Yu. Korchin}{The top-quark polarization}

%
\catchline{}{}{}{}{}
%

\title{The top-quark polarization beyond the Standard Model in electron-positron annihilation}

\author{I. V.~Truten$^{1,}$\footnote{i.truten@kipt.kharkov.ua}\; and  
A. Yu.~Korchin$^{1,2,}$\footnote{korchin@kipt.kharkov.ua} $^,$\footnote{Corresponding author}}

\address{$^1 $NSC ``Kharkov Institute of Physics and Technology'',  61108 Kharkiv, Ukraine\\
$^2$ V.N.~Karazin Kharkiv National University,  61022 Kharkiv, Ukraine}

\maketitle


\begin{abstract}
Electron-positron annihilation into a pair of top quarks is considered at the energy of the future collider CLIC. 
Polarization components of the top quark are calculated with the $\gamma t \bar{t}$ 
and $Z t \bar{t}$ interactions which follow from the Lagrangian of the effective field theory  
of the Standard Model. The polarization vector as a function of the scattering angle 
is calculated and averaged components of polarization are analyzed as functions of anomalous 
coupling constants. Extrema of these observables (maxima, minima and saddle points) are studied 
as functions of the $e^+ e^-$ energy.

\end{abstract}

\ccode{PACS numbers: 12.15.-y, 12.60.-i, 13.66.-a, 14.65.Ha}


\section{Introduction}

The top quark was discovered in 1994 at the Tevatron. 
Due to its enormous mass $m_t$=173.0 $\pm$ 0.4 GeV (direct measurements) \cite{Tanabashi:2018oca} 
the top quark is a short-lived particle so it can be detected by studying  
the products of the decay $t \to W^+ \, b \to \ell^+ \, \nu_\ell \, b$. 
The top quark is important in various models 
of ``new physics'' which is not described by the Standard Model (SM). 
The useful and detailed information about 
particle properties can be obtained from polarization experiments. 
In $pp$ collisions such measurements have been performed at the LHC at $\sqrt{s}=7$ TeV~\cite{Chatrchyan:2013wua, Aad:2013ksa}
and  $\sqrt{s}=8$ TeV~\cite{Khachatryan:2016xws} .   
  
The top-quark precision measurements at the $e^+e^-$ colliders are expected to be cleaner than 
those at the $pp$ colliders because interaction of electrons with the photon and Z-boson is well understood. 
Future experiments at the $e^+e^-$ colliders are supposed to be highly sensitive 
to physics beyond the Standard Model (BSM)~\cite{deBlas:2018mhx, Roloff:2018dqu}~. 
And the important direction in the progress of studying physics of the top quark is polarization 
measurements at the electron-positron colliders. 

The present paper discusses physical problems of interest for the future TeV scale linear accelerator -- Compact Linear Collider (CLIC).
It is expected that CLIC will improve measurements and limits on some physical parameters 
which have been obtained so far at the LHC and the Tevatron (see, {\it e.g.}, \cite{Vos:2016til, Garcia:2016}). 
CLIC is foreseen to operate with center-of-mass energies $\sqrt{s}$ =380 GeV, 1.5 TeV, 
and 3 TeV and with corresponding total luminosities of 500 fb$^{-1}$, 1.5 ab$^{-1}$, and 3 ab$^{-1}$, 
respectively~\cite{CLIC:2016zwp} .

In this work we study polarization of the top quark produced in the electron-positron annihilation, $e^+ \, e^- \to t  \, \bar{t}$. 
We calculate the top-quark polarization components in the framework of the effective field theory (EFT) Lagrangian of the SM which 
satisfies $SU(3)_C \times SU(2)_L \times U(1)_Y$ gauge symmetry. This Lagrangian contains the     
usual renormalizable SM Lagrangian ${\cal L}_{SM}^{(4)}$ 
and terms of dimension $d \ge 5$ which account for physics BSM. 
Terms of dimension six generate anomalous contributions to the photon-quark and Z-boson-quark 
interaction vertices which are similar in structure to radiative corrections in the SM.  

In Section \ref{subsec:lagrangian}, we consider the Lagrangian with the BSM 
coupling constants $\kappa$ and $\kappa_z$ which are directly related to the anomalous 
magnetic dipole moment (AMM) and anomalous weak magnetic dipole moment (AWMM), respectively. 
These couplings are expressed through the complex Wilson coefficients with
values constrained by the up-to-date fit in Ref.~\refcite{Buckley:2017} to experimental data 
from the Tevatron and the LHC Run 1 and Run 2. 
The authors \cite{Buckley:2017} analyzed cross sections at energies up to 13 TeV for 
the single and pair top production, and 
presented bounds on Wilson coefficients for the dimension-6 operators.  
Slightly different bounds have been suggested in Refs.~\citen{Bouzas:2013jha, Bouzas:2015} which used the 
electroweak data and $B$-meson decays. In our work the Wilson coefficients from 
\cite{Bouzas:2013jha, Bouzas:2015} are used for obtaining bounds on couplings $\kappa$ and $\kappa_z$. 

In Section \ref{subsec:polarization}, the polarization of 
the top quark is calculated in the reaction $e^+ \, e^- \to t \, \bar{t}$  at the center-of-mass 
energy $\sqrt{s}=380$ GeV.  Dependencies of the quark polarization on the anomalous 
couplings $\kappa$ and $\kappa_z$ are analyzed.
The cross section for unpolarized quarks and the components of polarization averaged over scattering angle 
 are studied as functions of couplings $\kappa$ and $\kappa_z$. 
The extreme points (maxima, minima and saddle points) of the corresponding two-dimensional surfaces  
are found and analyzed. A feasibility of measuring the BSM effects in polarization in future experiments 
is briefly estimated.  In Section \ref{sec:conclusions}, we present conclusions. 
\ref{app:A} and \ref{app:B} contain some details of formalism.


\section{\label{subsec:lagrangian} Lagrangian of $\gamma f \bar{f}$ and $Z f \bar{f}$ Interaction}

We consider the process of electron-positron annihilation into a pair of quarks 
\begin{equation}
e^-(k) + e^+(k^\prime) \to \left( \gamma^{*}, Z^{*} \right) \to t(p) + \bar{t}(p^\prime),
\label{eq:01}
\end{equation}
where $e^- (e^+)$ denotes an electron (positron) and $t (\bar{t})$ describes a top quark (antiquark) with 
the mass $m_t$. Here $p$ ($p^\prime$) is the top-quark (antiquark) four-momentum and $k$ ($k^\prime$) 
is the electron (positron) four-momentum.

We assume that the Lagrangian describing interactions of the photon and Z-boson with the quark 
includes the SM contribution and terms BSM which are related to new physics. 
For the photon the Lagrangian is
\begin{equation}
\mathcal{L}_{f\bar{f}\gamma}=e\bar{f} \left(Q_{f}\gamma^{\mu}A_{\mu}+ \frac{1}{4m_{f}}\sigma^{\mu\nu}
F_{\mu\nu}(\kappa+i\tilde{\kappa}\gamma_{5})\right)  f, 
\label{eq:02}
\end{equation}
and for the $Z$-boson the Lagrangian reads
\begin{eqnarray} \mathcal{L}_{f\bar{f}Z} &=& \frac{g}{2\cos{\theta_w}}\bar{f} \biggl( \gamma^{\mu}Z_{\mu} (v_f -a_f \gamma_5) 
+ \frac{1}{4m_{f}}\sigma^{\mu\nu}Z_{\mu\nu}(\kappa_z+i\tilde{\kappa}_z\gamma_{5})\biggr)  f .
\label{eq:03}
\end{eqnarray}
Here $f=(\ell, \, q)$ stands for the fermion field, $A^\mu$ and $Z^\mu$ are fields of 
the photon and $Z$-boson, 
\begin{eqnarray}
F_{\mu\nu} = \partial_{\mu}A_{\nu}-\partial_{\nu}A_{\mu}, \qquad
Z_{\mu\nu} = \partial_{\mu}Z_{\nu}-\partial_{\nu}Z_{\mu},
\label{eq:04}
\end{eqnarray}
and $g={e}/{\sin{\theta_w}}$ with $\theta_w$ denoting the weak mixing angle.

In these equations $Q_f$ is electric charge of the fermion in units of the positron 
charge $e=\sqrt{4 \pi \alpha_{em}}$,  \  $\alpha_{em}$ is the fine structure constant, 
$v_f, \, a_f$ are the vector- and axial-vector couplings 
\begin{eqnarray}
v_f = T^f_3-2 Q_f \sin^2{\theta_w}, \qquad
a_f = T^f_3.
\label{eq:05}
\end{eqnarray}
with $T^f_3$ being the $3^{\rm rd}$ component of the weak isospin. In particular, $v_e=-\frac{1}{2}+2 \sin^2{\theta_w}, \; a_e=-\frac{1}{2}$ 
for an electron, and $v_t =\frac{1}{2}-\frac{4}{3} \sin^2{\theta_w}, \; a_t=\frac{1}{2}$ for a top quark. 

The terms in Eqs.~(\ref{eq:02}) and (\ref{eq:03}) proportional to $\kappa$ and $\kappa_z$ 
determine the $CP$-even interaction, while those proportional to $\tilde{\kappa}$ and $\tilde{\kappa}_z$ 
determine the $CP$-odd interaction.

The BSM terms in Eqs.~(\ref{eq:02}) and (\ref{eq:03}) for $f=q$ arise in the framework of the EFT. 
These terms in the EFT Lagrangian appear after integration over heavy degrees of freedom connected with 
new physics. The EFT Lagrangian has the structure \cite{Burges:1983, Leung:1984, Buchmuller:1985} 
\begin{equation}
\mathcal{L}_{eff}= 
\mathcal{L}_{SM}^{(4)} +  \frac{1}{\Lambda} \sum_{i} C_i^{(5)} \mathcal{O}_i^{(5)} 
+ \frac{1}{\Lambda^2} \sum_{i} C_i^{(6)} \mathcal{O}_i^{(6)} + \ldots ,
\label{eq:06}
\end{equation}
where $\Lambda$ is the scale of new physics,  gauge-invariant operators 
$\mathcal{O}_i^{(d)}$ for $d \ge 5$ are 
expressed through the fields of the SM, and $C_i^{(d)}$ are Wilson coefficients. 

As shown in Refs.~\citen{AguilarSaavedra:2008, Grzadkowski:2010}, among 59 operators of 
dimension six there exist only two relevant for the $\gamma t \bar{t}$ and $Z t \bar{t}$ 
interaction operators (the superscript $(6)$ is omitted hereafter):
\begin{eqnarray}
\mathcal{O}_{u B \phi}^{33} &=& \bar{q}_{L 3 } \sigma^{\mu \nu} t_R \tilde{\phi}  B_{\mu \nu} + h.c., \nonumber \\
\mathcal{O}_{u W }^{33} &=& \bar{q}_{L 3 } \sigma^{\mu \nu} \tau^a t_R \tilde{\phi}  W_{\mu \nu}^a + h.c. ,
\label{eq:07}
\end{eqnarray}
where index `3' means the $3^{\rm rd}$ quark generation, $\tilde{\phi} = i \tau_2 \phi^\star$, \ $\phi$ is the SM Higgs doublet, 
and strength tensors of the gauge fields are defined as
\begin{eqnarray}
B_{\mu \nu} = \partial_\mu B_\nu - \partial_\nu B_\mu, \qquad
W_{\mu \nu}^a = \partial_\mu W_\nu^a - \partial_\nu W_\mu^a - g \varepsilon^{abc} W_\mu^b W_\nu^c,
\label{eq:08}
\end{eqnarray}
where $a, \, b, \, c=1,\, 2, \, 3$.
Eqs.~(\ref{eq:07}), after replacing 
$\langle \tilde{\phi} \rangle \to \frac{1}{\sqrt{2}} 
\begin{pmatrix} v \\ 0 \end{pmatrix}$,
give rise to the couplings $\kappa$, $\tilde{\kappa}$ and 
$\kappa_z$, $\tilde{\kappa}_z$ in Eqs.~(\ref{eq:02}) and (\ref{eq:03}). 

For the electromagnetic interaction (\ref{eq:02}) one finds from (\ref{eq:07}) the $\gamma t \bar{t} $ couplings 
\begin{eqnarray}
 \kappa &=& \sqrt{2} \frac{m_t}{m_z} \, \frac{1}{c_w s_w}  \mathrm{Re} \, 
(s_w \bar{C}_{uW}^{33}+c_w \bar{C}_{uB \phi}^{33}),\nonumber \\  
 \tilde{\kappa} &=& \sqrt{2} \frac{m_t}{m_z} \, \frac{1}{c_w s_w} \mathrm{Im} \, 
(s_w \bar{C}_{uW}^{33} +c_w \bar{C}_{uB \phi}^{33} ),
\label{eq:09}
\end{eqnarray}
which are related to the anomalous magnetic and electric dipole moments of the top quark, respectively. 
Hereafter $c_w \equiv \cos{\theta_w}, \, s_w \equiv \sin{\theta_w}$, 
$m_z$ is the mass of the $Z$ boson and we defined 
\begin{eqnarray}
\bar{C}_{uW}^{33} \equiv \frac{v^2}{\Lambda^2}{C}_{uW}^{33}, \qquad
\bar{C}_{uB \phi}^{33} \equiv \frac{v^2}{\Lambda^2} {C}_{uB \phi}^{33}, 
\label{eq:10}
\end{eqnarray} 
where $v=246$ GeV is the vacuum value of the scalar Higgs 
field and the scale parameter $\Lambda$ is taken to be 1 TeV \cite{Bouzas:2015} .

For the Z-boson interaction (\ref{eq:03}) one finds from (\ref{eq:07}) the relations for the $Z t \bar{t}$ couplings 
\begin{eqnarray}
\kappa_z &=& 2\sqrt{2} \frac{m_t}{m_z} \mathrm{Re} \, (c_w \bar{C}_{u W}^{33} -s_w \bar{C}_{u B \phi}^{33}), \nonumber \\
\tilde{\kappa}_z &=& 2 \sqrt{2} \frac{m_t}{m_z} \,   \mathrm{Im} \, (c_w \bar{C}_{u W}^{33}-s_w \bar{C}_{u B \phi}^{33}),
\label{eq:11}
\end{eqnarray}
which are related to the anomalous weak magnetic and weak electric dipole moments, respectively. 

All the information on the anomalous $\gamma t \bar{t}$ and $Z t \bar{t}$ interactions is contained in two 
complex coefficients $\bar{C}_{u W}^{33}$ and $\bar{C}_{u B \phi}^{33}$. 
The bound on the coefficient $\bar{C}_{u B \phi}^{33}$ comes from experiments on the decay of $B$ mesons,
$B \to X_s \gamma$, in particular, on the measured branching ratio and $CP$ asymmetry. 
The coefficient $\bar{C}_{u W}^{33}$ is constrained by the LHC data at 7 and 8 TeV on the 
single top-quark production and $W$ polarization in the top-quark decay.  
According to \cite{Bouzas:2013jha, Bouzas:2015} the present bounds on the real values of 
the coefficients, in our definition (\ref{eq:10}), are
\begin{eqnarray}
\bar{C}_{u W}^{33} = (-0.061, \, 0.030), \qquad
\bar{C}_{u B \phi}^{33} = (-0.363, \, 0.054).
\label{eq:12}
\end{eqnarray}
The imaginary part of the coefficients is not constrained from the data and is chosen zero in our calculations. 
Correspondingly the couplings $\tilde{\kappa} $ and $\tilde{\kappa}_z$ are absent. 

Substituting the bounds in (\ref{eq:12}) in Eqs.~(\ref{eq:09}), (\ref{eq:11}) one obtains 
the bounds on the coupling constants of the BSM physics:
\begin{eqnarray}
\kappa = (-2.21, \, 0.40), \qquad
\kappa_z = (-0.43, \, 1.08).
\label{eq:13}
\end{eqnarray}
In the following sections the values (\ref{eq:13}) will be used.

In the considered process $e^+ \, e^- \to t \, \bar{t}$ the invariant energy squared $s  \geq 4 m_t^2 $,  
and the threshold is far from the photon pole ($s =0$) and $Z$-boson pole ($s=m_z^2$). 
Therefore it is appropriate to introduce the $q^2$-dependence of the couplings 
$\kappa$ and $\kappa_z$ as follows  
\begin{eqnarray}
 \kappa (s) = Q_t F^\gamma_2 (s),  \qquad
\kappa_z (s) = v_t F^Z_2 (s),
 \label{eq:14}
\end{eqnarray}
where $F^{\gamma, \, Z}_2 (s)$ are some form factors. 

The coupling $\kappa (0)$ is directly related to the anomalous part of the quark 
magnetic moment $\mu_t$, via the relations \cite{Hollik:1998vz}
\begin{eqnarray}
\kappa (0) = Q_t F^\gamma_2 (0) = Q_t a_t, \qquad
\mu_t = \frac{e Q_t }{2 m_t} (1+ a_t).  
\label{eq:15}
\end{eqnarray}

In the SM, for a pointlike Dirac particle without radiative corrections, $a_t$ and $\kappa (0) $ are equal to zero. 
Modification arises due to radiative corrections and can also stem from new physics. 
The radiative corrections to $a_t$ have been evaluated 
to one loop in the EW theory~\cite{Czarnecki:1995sz, Hollik:1988ii, Bernabeu:1997je} and two loops 
in QCD~\cite{Bernreuther:2004ih, Bernreuther:2004th, Bernreuther:2005rw} . 
Here we cite the results of Ref.~\refcite{Bernreuther:2005gq} for radiative corrections to the couplings 
$\kappa (0)$ and $\kappa_z (0)$ 
which are evaluated to the two loops in QCD and to the lowest order in electroweak couplings 
\begin{eqnarray}
\kappa(0)_{rc} &=& Q_t  F^{\gamma}_2(0)_{rc}  = 2.0  \times 10^{-2}, \nonumber \\  
\kappa_z(0)_{rc}  &=& v_t  F^{Z}_2(0)_{rc} =  5.75 \times 10^{-3}, 
\label{eq:16}
\end{eqnarray}
where
\begin{eqnarray}
F^{\gamma}_2(0)_{rc}&=& F^{Z}_2(0)_{rc} \equiv F_2 (0)_{rc} =  3.0  \times 10^{-2}.
\label{eq:17}
\end{eqnarray}   
Here `rc'  is abbreviation for `radiative corrections' and $F_2(s)_{rc}$ is the form factor  defined in~\cite{Bernreuther:2005gq}.  
The values in (\ref{eq:16}) are calculated \cite{Bernreuther:2005gq} with the strong coupling $\alpha_s (\mu)=0.108$ 
at the renormalization scale $\mu=m_t$. 
The QED correction to $\kappa (0)$ in the leading order, $ \kappa (0)_{rc, \, QED} =Q_t^3 \alpha_{em} / (2 \pi)$, turns out 
to be of the same order as the three-loop correction in QCD and is much less than the value in Eqs.~(\ref{eq:16}).

The bounds in Eqs.~(\ref{eq:13}) are much wider than the values in (\ref{eq:16}). 
It will be important to obtain the more precise constraints on $\kappa (s)$ and $\kappa_z(s)$ 
from high-precision measurements at $e^+e^-$ colliders. 


\section{\label{subsec:polarization} Top-Quark Polarization and Discussion of Results}

\begin{figure}[tbh]
\centerline{\includegraphics{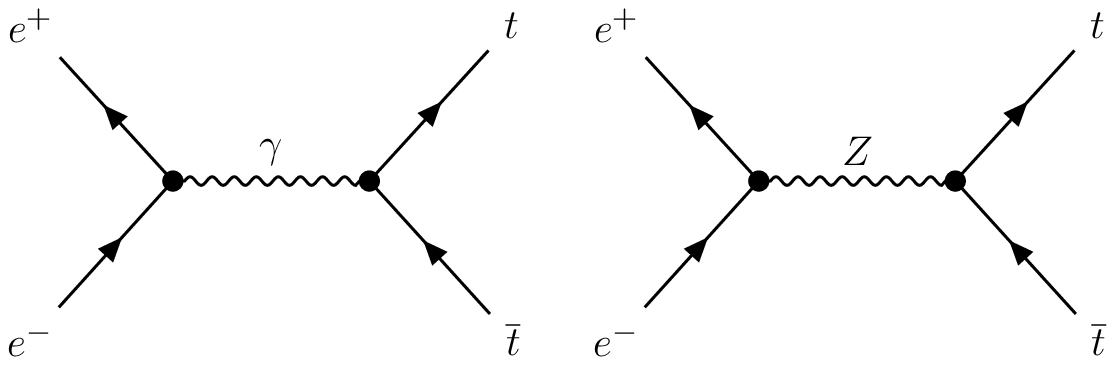}}
\caption{Tree-level diagrams for the process 
$e^- \, e^+ \to \left( \gamma^{*}, Z^{*} \right) \to t \, \bar{t}$.}
\label{fig:1}
\end{figure}

In the ``tree level'' approximation this process is described by two diagrams with the photon and $Z$-boson 
exchanges in the $s$-channel (see Fig.~\ref{fig:1}). The matrix element of this process is presented in~\ref{app:A}.

Since outgoing quarks are polarized, we use the covariant fermion density matrix:
\begin{eqnarray}
u(p,m)\bar{u}(p,m) &=& \tfrac{1}{2}(\not{p}+m)  ({1 + \gamma^5 \vslash{a}   }), \nonumber \\
v(p,m)\bar{v}(p,m) &=& \tfrac{1}{2}(\not{p}-m) ({1 + \gamma^5 \vslash{a}}),
\label{eq:18}
\end{eqnarray}
where $u, \bar{u}$ are the Dirac spinors for a particle, $v, \bar{v}$ -- for an antiparticle.
In Eqs.~(\ref{eq:18})  $\vslash{a}\equiv \gamma_\mu a^\mu$, where $a^\mu$ is the four-vector of fermion polarization. 
In the rest frame of fermion, $p=(m, \, \vec{0})$, this vector can be expressed through the three-vector $\vec{\xi}$ of 
the fermion double average spin
\begin{equation}
a=\left(0,  \, \vec{\xi}\right),         \qquad a \cdot p=0.
\label{eq:19}
\end{equation}
Similarly the polarization vector of the antiquark, $a^\prime$, can be defined.

\begin{figure}[tbh]
\centerline{\includegraphics[scale=0.85]{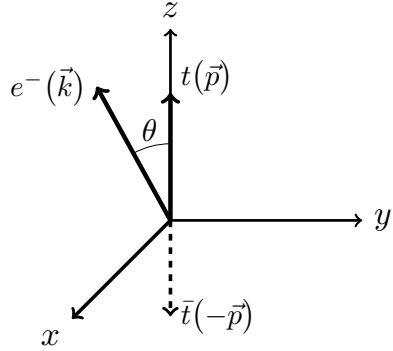}}
\caption{Coordinate system.}
\label{fig:2}
\end{figure}

Further the coordinate system is chosen in the same way as in Ref.~\refcite{Arens:1992} and is 
shown in Fig.~\ref{fig:2}. 
The quark momenta are directed along the OZ axis and the lepton momenta lie in the XOZ plane, 
$\theta$ is the angle between the electron and the quark momenta.

In the center-of-mass system the four-momenta of electron and positron are 
\begin{eqnarray}
k &=& \bigl(E, \, E \sin{\theta}, \, 0, \, E \cos{\theta}\bigr),  \nonumber \\ 
k^\prime &=& \bigl(E, \, -E \sin{\theta}, \, 0, \, -E \cos{\theta}\bigr), 
\label{eq:20}
\end{eqnarray}
and the momenta and polarization vectors of quarks and antiquarks are
\begin{eqnarray}
p &=& \bigl(E, \, 0, \, 0, \, V E\bigr), \nonumber \\ 
p^\prime &=& \bigl(E, \, 0, \, 0, \, -V E\bigr),  \nonumber \\
a &=& \bigl(\gamma V \xi_z, \, \xi_x,  \, \xi_y, \, \gamma \xi_z\bigr),  \nonumber \\ 
a^\prime &=& \bigl(-\gamma V \xi^{\prime}_z, \, \xi^{\prime}_x, \, \xi^{\prime}_y, \, \gamma \xi^{\prime}_z\bigr),
\label{eq:21}
\end{eqnarray}
where $E= {\sqrt{s}}/{2}$ is the energy of the lepton or quark, $V$ -- velocity of the 
top quark, $\gamma={E}/{m_t}=1/ \sqrt{1-V^2}$ is the Lorentz factor.
Eqs.~(\ref{eq:21}) are obtained by using Lorentz transformation from the rest frame of the quarks.

The Lagrangians (\ref{eq:02}) and (\ref{eq:03}) generate the BSM contribution  
to the matrix element squared, $|M|^2 = |M^{SM}+M^{BSM}|^2= |M^{SM}|^2 + 2 \mathrm{Re} (M^{SM *} \, M^{BSM} ) +|M^{BSM}|^2$.  
The matrix element $M^{BSM}$ is \ ${\cal O}(\Lambda^{-2})$ in EFT and strictly speaking 
the term $|M^{BSM}|^2$ can be omitted. However, our explicit calculation shows that neglecting 
this term, for some values of $\kappa$ and $\kappa_z$, can lead to negative cross sections. 
In view of this we keep all terms in $|M|^2$ in calculation of the cross section.   

For polarized quarks the cross section is a function of the polarization components 
$\xi_i, \, \xi^\prime_j$ \ ($i,\, j \, = \, (x,y,z)$) and has the form 
\begin{equation}
\frac{d\sigma}{d\Omega}(\theta, \vec{\xi}, \vec{\xi^\prime}) = 
\frac{3 \alpha^2_{em}(s) V}{16 E^2} N \, \bigl(1 + {P_i}{\xi _i} +P_{j}^{\prime} \xi^\prime_{j}+Q_{ij} \xi_{i} \xi_{j}^{\prime}\bigr),
\label{eq:22}
\end{equation}
where $N$ is related to the cross section for unpolarized quarks as
\begin{equation}
\frac{3\alpha^2_{em}(s) V}{16 E^2} N = \frac{1}{4} \frac{d\sigma_0}{d\Omega}(\theta). 
\label{eq:23}
\end{equation}
The factor 3 accounts for the quark colors, $d \Omega$ is the solid angle for the top quark, 
$\alpha_{em}(s)$ is the energy-dependent fine structure constant, 
$\vec{P}$ and $\vec{P}^{\prime}$ are the quark and antiquark polarizations, 
respectively, which arise in the process of the quark production. 

Here, we do not consider correlations of polarizations described by the term $\sim Q_{ij}$, 
and neglect the last term in Eq.~(\ref{eq:22}). Therefore  
\begin{equation}
\frac{d\sigma}{d\Omega}(\theta, \vec{\xi}, \vec{\xi^\prime}) = 
\frac{1}{4} \frac{d\sigma_0}{d\Omega} (\theta) \,  \bigl(1 + {P_i}{\xi _i} +P_{j}^{\prime} \xi_{j}^{\prime}\bigr).
\label{eq:24}
\end{equation}

\begin{figure*}[tbh]
\begin{minipage}[h]{0.47 \linewidth}
\centerline{\includegraphics[scale=0.54]{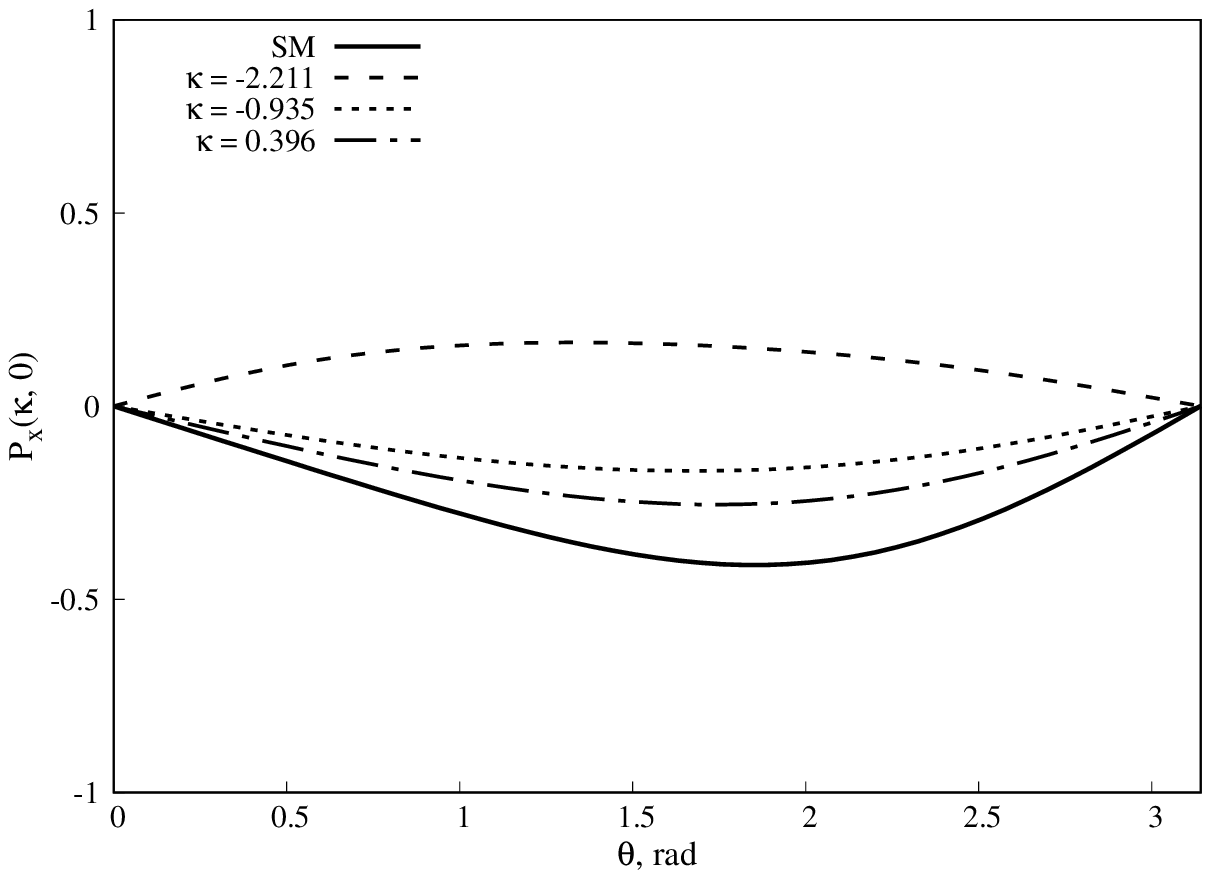}}
\end{minipage}
\hfill
\begin{minipage}[h]{0.47 \linewidth}
\centerline{\includegraphics[scale=0.54]{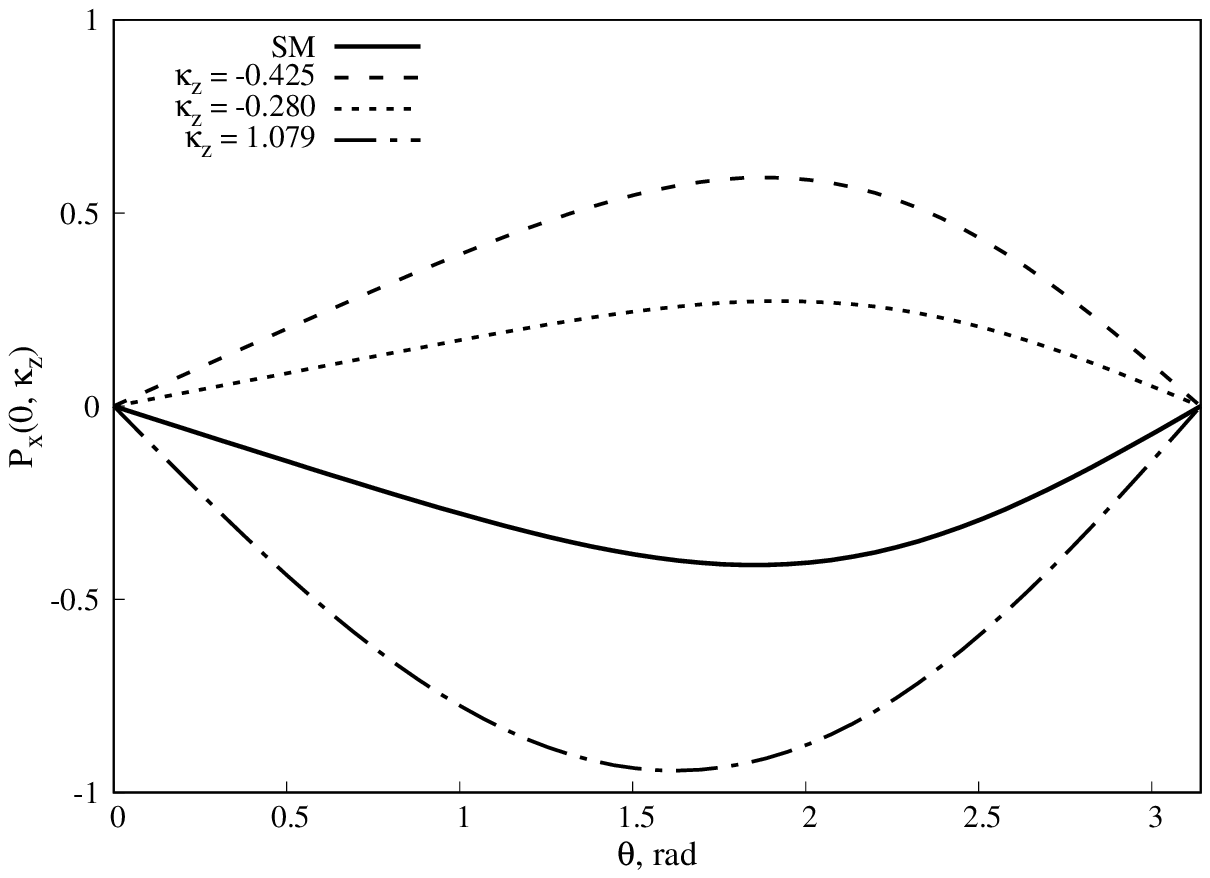}}
\end{minipage}
\vfill
\begin{minipage}[h]{0.47 \linewidth}
\centerline{\includegraphics[scale=0.54]{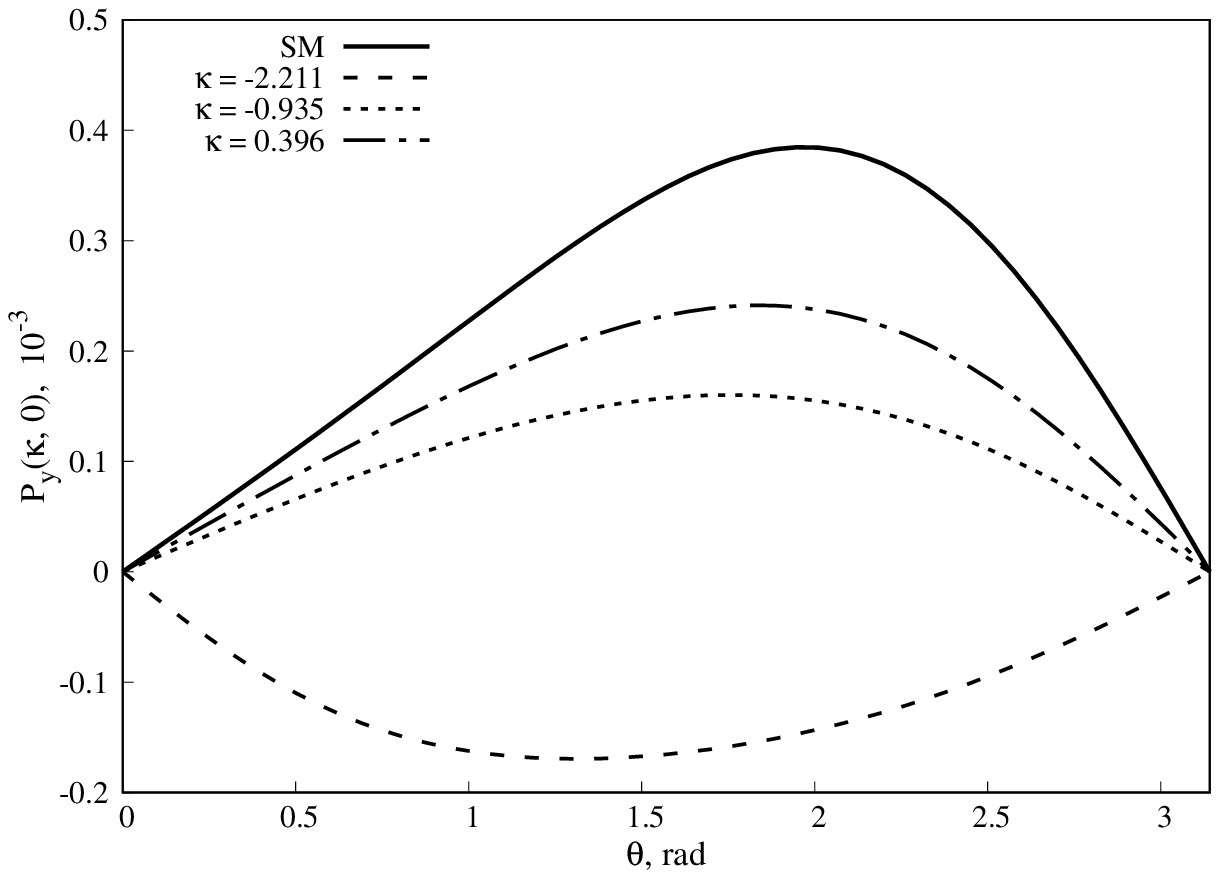}}
\end{minipage}
\hfill
\begin{minipage}[h]{0.47 \linewidth}
\centerline{\includegraphics[scale=0.54]{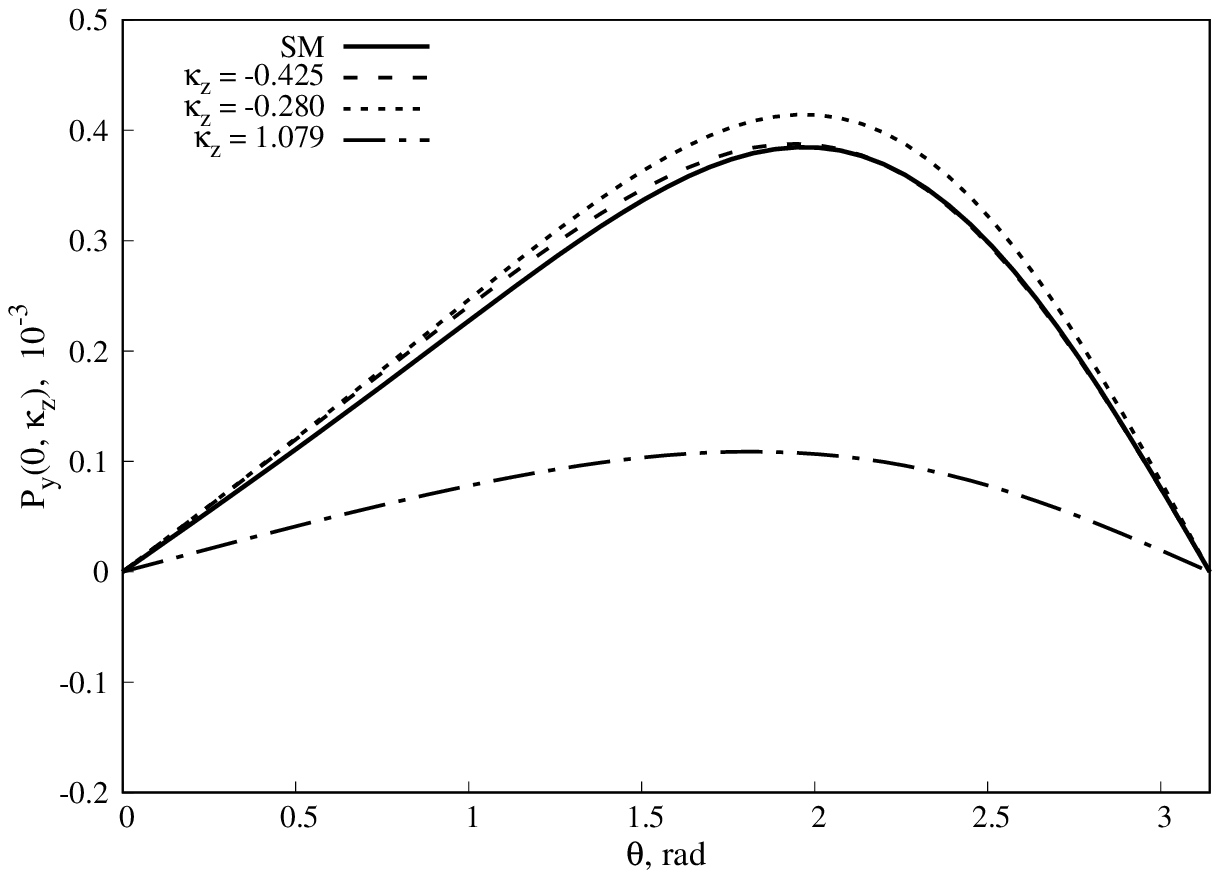}}
\vfill
\end{minipage}
\vfill
\begin{minipage}[h]{0.47 \linewidth}
\centerline{\includegraphics[scale=0.54]{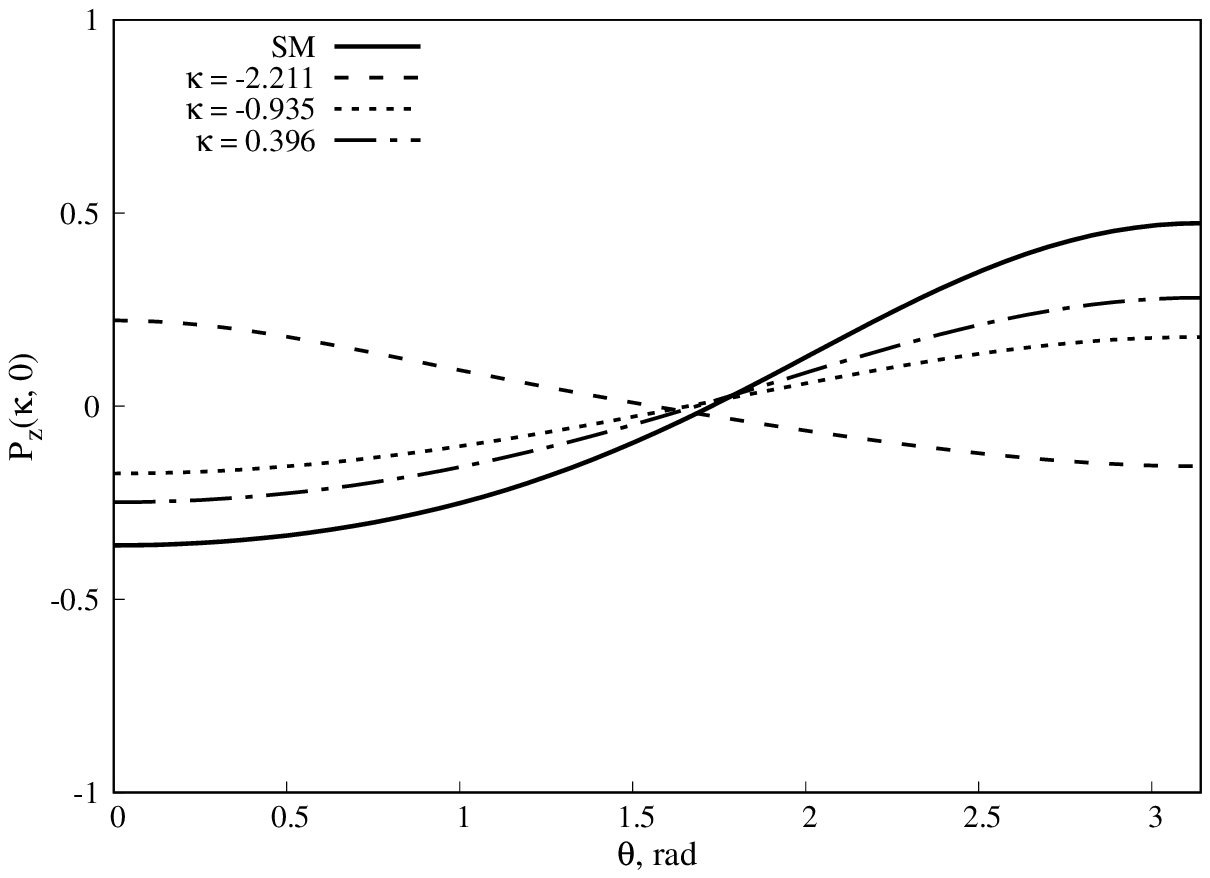}}
\end{minipage}
\hfill
\begin{minipage}[h]{0.47 \linewidth}
\centerline{\includegraphics[scale=0.54]{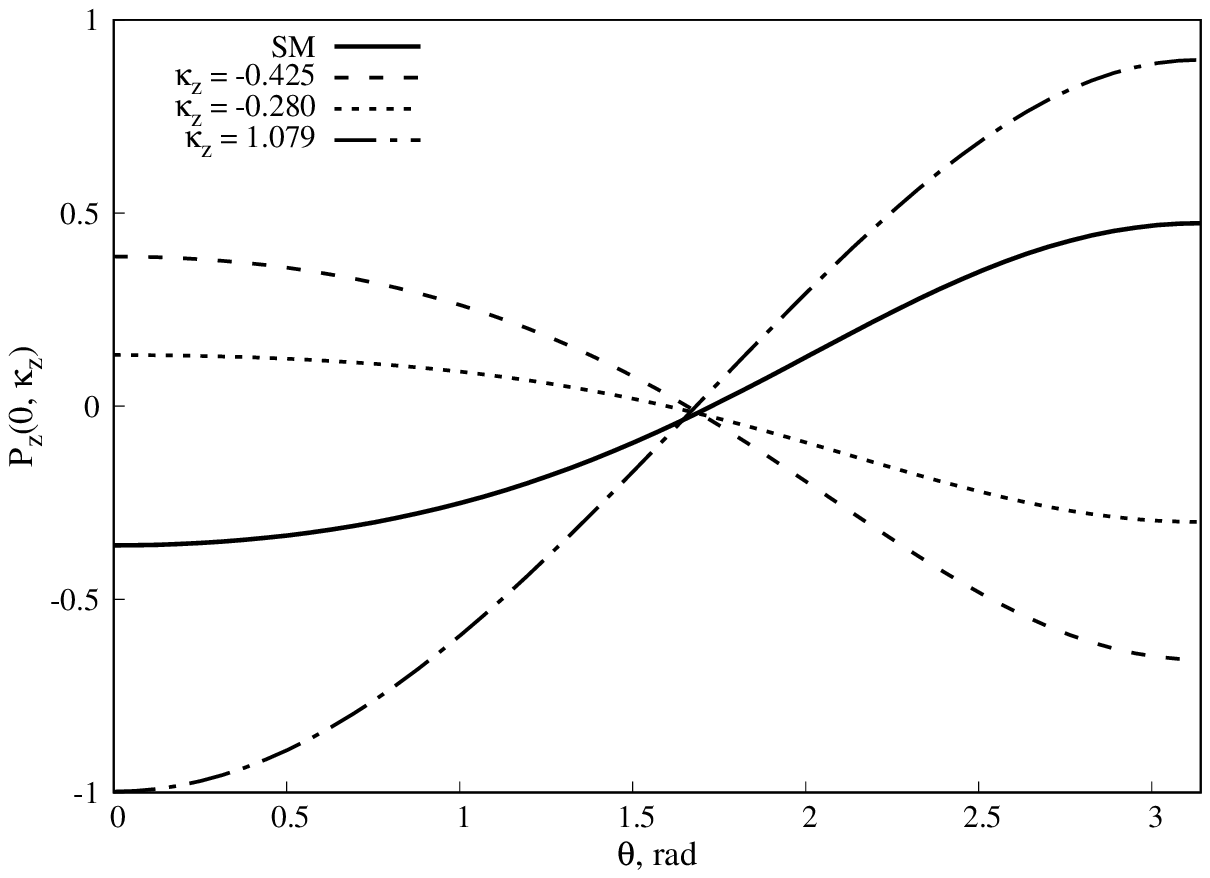}}
\vfill
\end{minipage}
\vfill
\caption{Components of the top-quark polarization as functions of the scattering 
angle $\theta$ for a center-of-mass energy $\sqrt{s}=380$ GeV. 
Left panel shows polarization components for several values of the BSM coupling 
$\kappa$ while taking $\kappa_z=0$, and the right panel shows polarization components 
for several values of the BSM coupling $\kappa_z$ while taking $\kappa=0$.}
\label{fig:3}
\end{figure*}
The components of the top-quark polarization $P_i (\kappa, \kappa_z)$ are given in \ref{app:A} and  
shown in Fig.~\ref{fig:3}. 
Calculations are performed for the invariant energy $\sqrt{s}=380$ GeV and are obtained using the bounds 
(\ref{eq:13}) for the couplings $\kappa$ and $\kappa_z$. 
Besides we use numerical values of physical constants in Table~\ref{tab:1}.
\begin{table}[tph]
\tbl{Numerical values of the physical constants~\cite{Tanabashi:2018oca}}
{\begin{tabular}{@{}cccc@{}} \toprule
Constant & Value & Constant & Value \\
\colrule
$m_t$ & 173.0 GeV & $a_e$ & $-0.50123$ \\
$m_z$ & 91.1876(21) GeV & $a_t$ & 0.5 \\
$\Gamma_z$ & 2.4952(23) GeV & $v_e$ & $-0.03783$ \\
$\sin^2{\theta_w}$ & 0.23155(4) & $v_t$ & 0.191\\ \botrule
\end{tabular} \label{tab:1}}
\end{table}

The solid curves in Fig.~\ref{fig:3} show calculations in the SM, in which the coupling constants in 
(\ref{eq:09}) and (\ref{eq:11}) are equal to zero. 
The other curves show calculations with the BSM coupling constants $\kappa$ (the plots on the left) 
and $\kappa_z$ (the plots on the right) which are constrained in Eqs.~(\ref{eq:13}).

First, we note that the component $P_y$ of the polarization vector, which is 
perpendicular to the reaction plane, is very small in the SM and in the considered extension of the SM. 
This component is generated at tree level from the $\gamma - Z$ interference and is proportional to the imaginary part of 
the $Z$-boson propagator. $P_y$ is smaller by several orders of magnitude than the other components 
of the polarization and takes values of the order $10^{-4} - 10^{-3}$.   
 
Second, the transverse to the quark momentum component, $P_x$, is sensitive to the coupling $\kappa$, 
and especially to the coupling $\kappa_z$ (see upper plots in Fig.~\ref{fig:3}). Variation of the coupling $\kappa_z$ can 
change considerably the polarization $P_x$. 

Third, the longitudinal component of the polarization vector $P_z$ is sizable in the SM, 
and it is also sensitive to the BSM couplings. It is interesting to note that the polarization $P_z$ reaches value +1 
for $\kappa_z=0.566$ and $\kappa=0$ at the forward angle, correspondingly the transverse polarization vanishes 
at this angle.     
    
Based on Fig.~\ref{fig:3} one can see that the terms BSM 
make quite a large contribution to the polarization of the top quark. 

\begin{figure*}[tbh]
\vskip1mm
\begin{minipage}[h]{0.47 \linewidth}
\centerline{\includegraphics[scale=0.55]{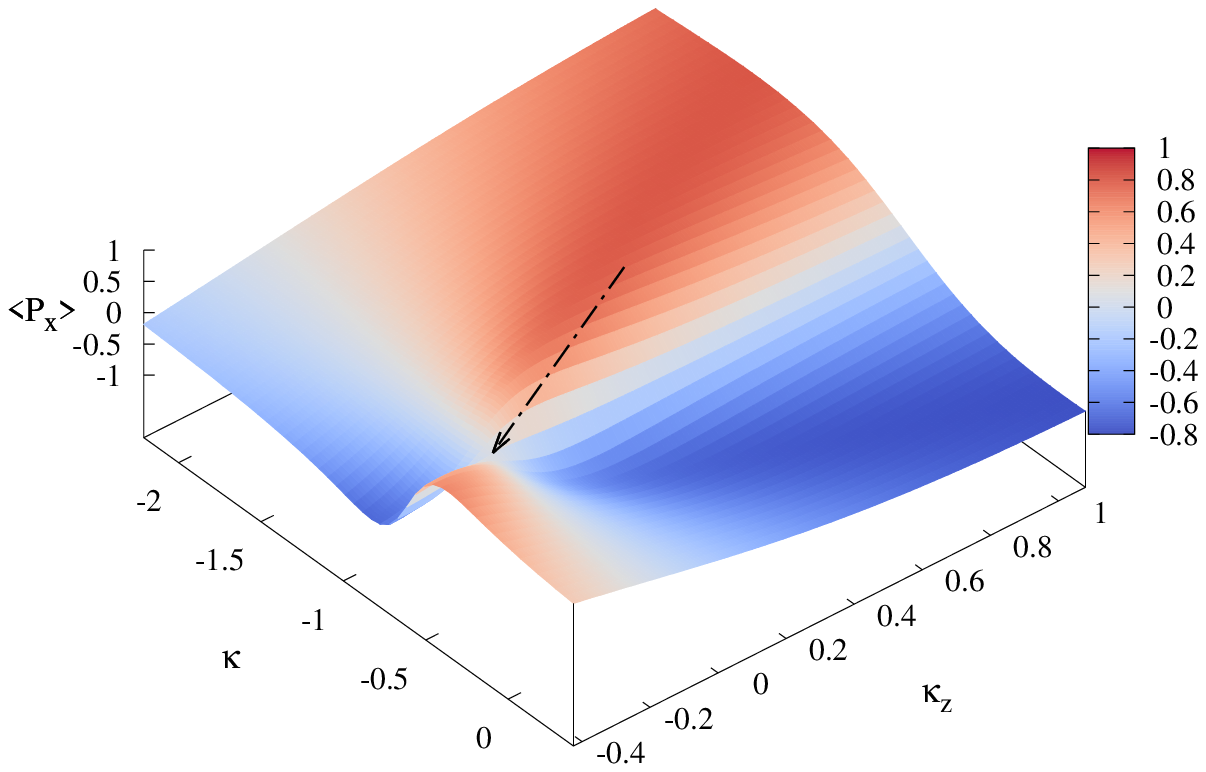}}
\end{minipage}
\hfill
\begin{minipage}[h]{0.47 \linewidth}
\centerline{\includegraphics[scale=0.55]{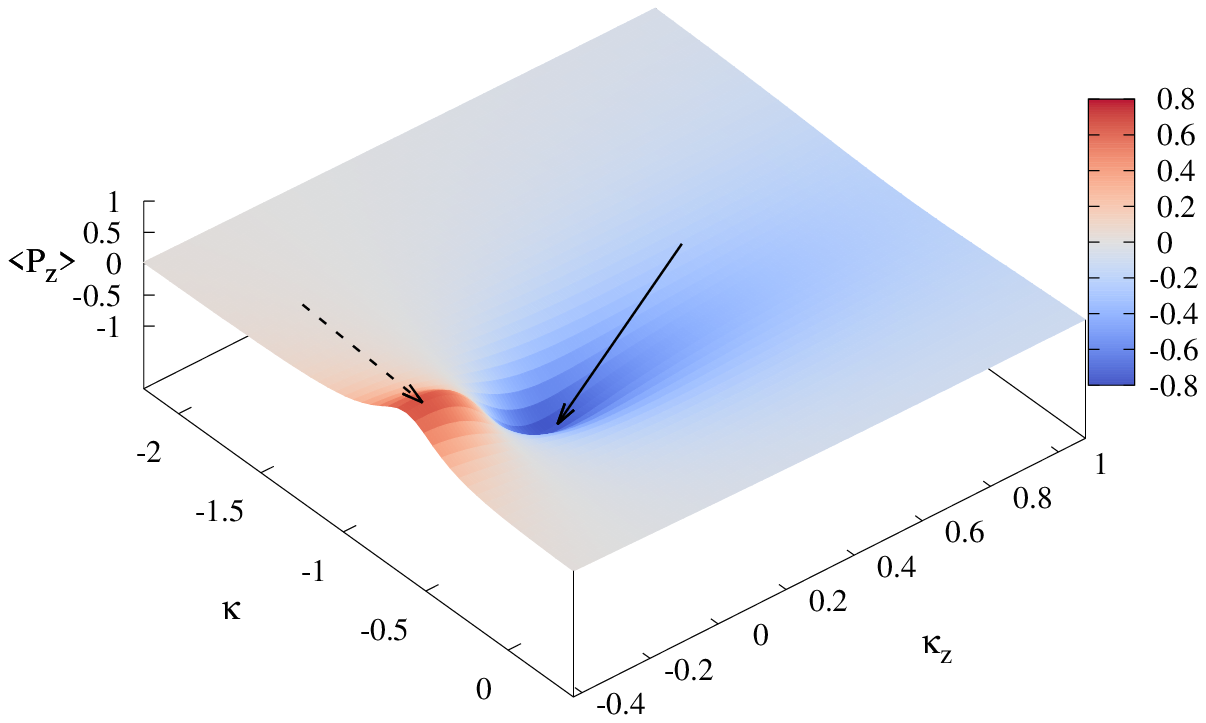}}
\end{minipage}
\vfill
\begin{minipage}[h]{0.47 \linewidth}
\centerline{\includegraphics[scale=0.55]{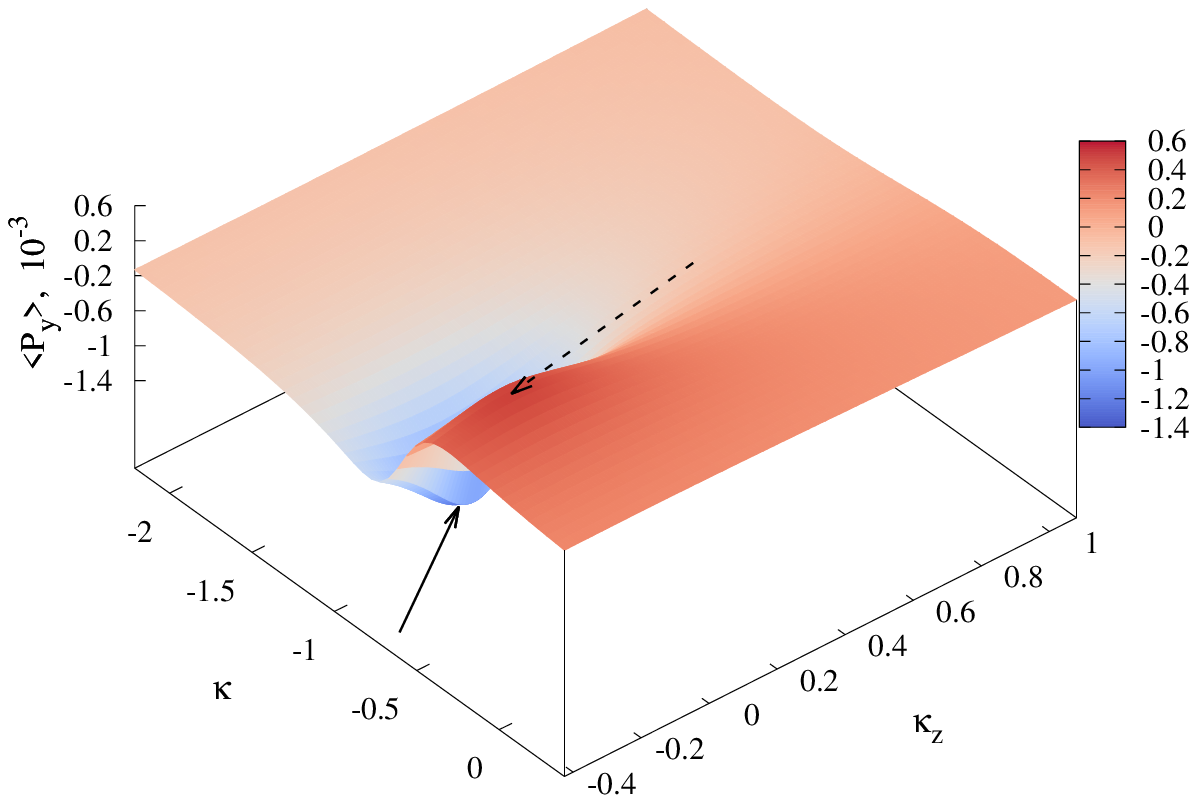}}
\end{minipage}
\hfill
\begin{minipage}[h]{0.47 \linewidth}
\centerline{\includegraphics[scale=0.55]{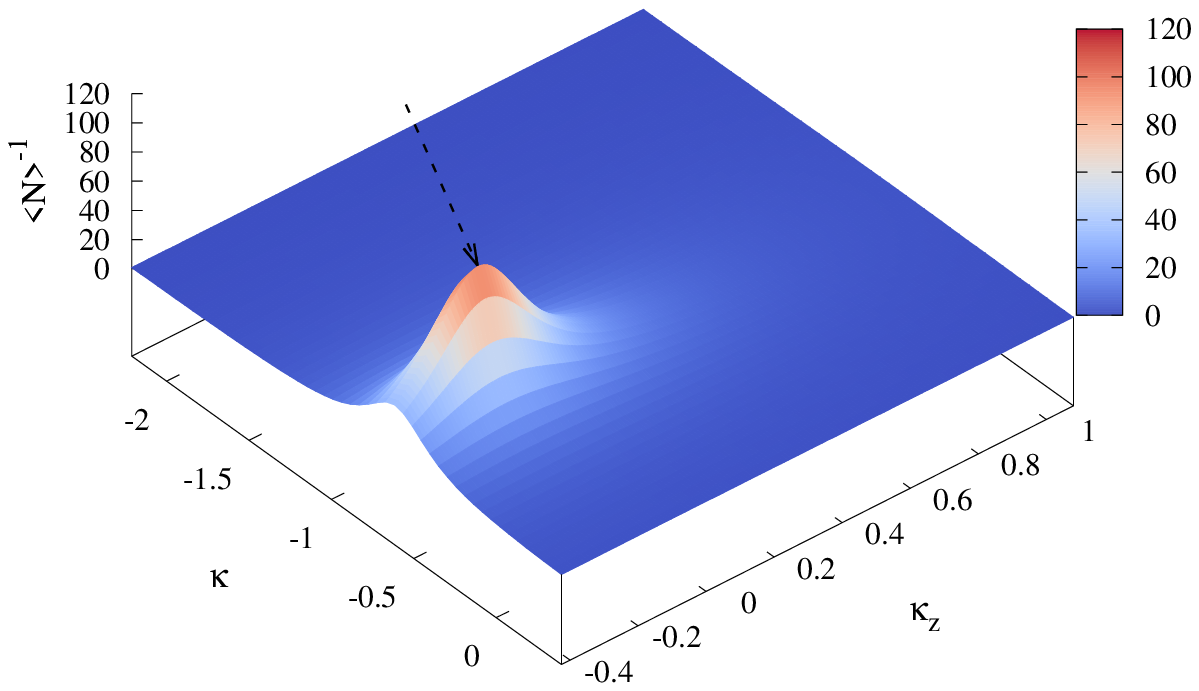}}
\end{minipage}
\vfill
\vskip-3mm
\caption{The averaged polarizations: transverse $\langle P_x \rangle $, perpendicular to scattering plane 
$\langle P_y \rangle $, longitudinal $\langle P_z \rangle $, and the quantity 
${\langle N \rangle}^{-1} $, versus the coupling constants $\kappa$ and $\kappa_z$. 
The center-of-mass energy is $\sqrt{s}=380$ GeV. Dashed arrows indicate the maximum 
value, solid arrows -- the minimum value, and dashed-dotted arrow -- the saddle point.}
\label{fig:4}
\end{figure*}

In order to demonstrate the simultaneous effect of both couplings $\kappa$ and $\kappa_z$ 
on the polarization components we perform averaging over the scattering angle in the following way   
\begin{equation}
\langle P_i \rangle =  \frac{\langle N P_i  \rangle}{\langle N \rangle},
\label{eq:25}
\end{equation}
for $i=(x,y,z)$. Here we define the average value of any quantity dependent on the scattering angle as
\begin{equation}
\langle A \rangle \equiv  \frac{1}{2} \int_{0}^{\pi} A(\theta) \, \sin \theta \, d \theta .
\label{eq:26}
 \end{equation}

The polarization components averaged in this way for the energy $\sqrt{s}= 380$ GeV are shown in Fig.~\ref{fig:4} 
(see~\ref{app:B} for explicit expressions).
As can be seen from Fig.~\ref{fig:4}, 
the surfaces $\langle P_i \rangle $ as functions of $\kappa$ and $\kappa_z$ have extreme points: maxima, minima and saddle points.  
The averaged transverse component $\langle P_x \rangle $ has a saddle point at the values $\kappa=-0.609$,  \ $\kappa_z=-0.177$.
The perpendicular to the scattering plane component $\langle P_y \rangle $ and the longitudinal component $\langle P_z \rangle $ have 
the sharp minimum and also maximum. The component $\langle P_y \rangle $ gives little information being extremely small and not 
significant from experimental point of view.
The minimal value of $\langle P_z \rangle $ is $-0.7843$ which appears at 
 $\kappa=-0.6237$ and $\kappa_z=-0.014$, while the maximum value, $0.6523$, appears at $\kappa=-0.623$ and $\kappa_z=-0.377$. 

Let us analyze the maximum of ${\langle N \rangle}^{-1} $ proportional to the inverse cross section for the unpolarized quarks 
(Eq.~(\ref{eq:B4}) in~\ref{app:B}).  The surface ${\langle N \rangle}^{-1} $ has a distinct maximum at $\kappa=-0.624, \ \kappa_z=-0.179$ with 
the value ${\langle N \rangle}^{-1}_{max} = 102.32$.
If we calculate the ratio $\frac{\kappa (s)}{\kappa_z(s)} = 3.48$, then it is seen that this ratio is the same as   
the ratio of the couplings $\kappa$ and $\kappa_z$ in the SM with radiative corrections (see Eqs.~(\ref{eq:16}))   
\begin{eqnarray}
 \frac{\kappa(0)_{rc} }{\kappa_{z}(0)_{rc}}=\frac{Q_t}{v_t} &\cong& 3.48.
 \label{eq:27}
\end{eqnarray}
Note that the ratio (\ref{eq:27}) is valid for any upper quark $u, \, c, \, t$. 
For the lower quarks $d, \, s, \, b \, $ the corresponding ratio is very different, $\frac{Q_q}{v_q}=0.96$,  
due to the different electric charges and weak isospin values.

\begin{figure}[tbh]
\centerline{\includegraphics[scale=0.75]{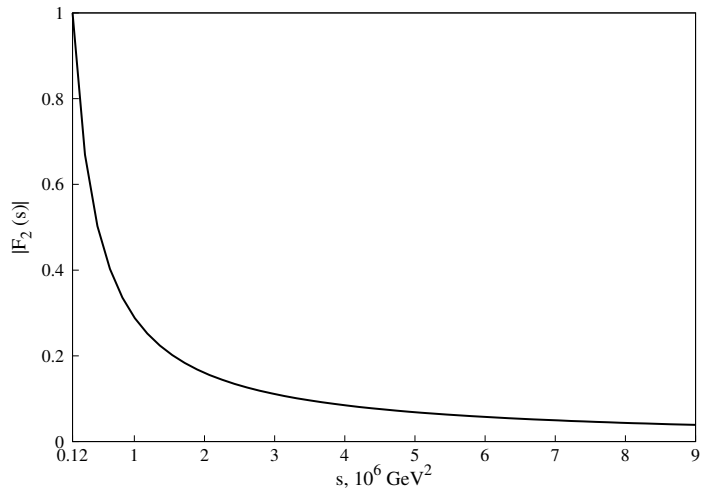}}
\caption{The form factor modulo as a function of the invariant energy squared $s$.  }
\label{fig:5}
\end{figure}

The location of the maximum of the surface ${\langle N \rangle}^{-1}$ changes with changing the  
energy $\sqrt{s}$.  We can make an assumption, albeit without proof, that the true values of the anomalous 
couplings $\kappa(s)$ and $\kappa_z (s)$ correspond to the position of the maximum of ${\langle N \rangle}^{-1}$, 
or minimum of the cross section for unpolarized quarks.  
Then this allows one to find variation of form factors in (\ref{eq:14}) with the energy. 
The photon and Z-boson form factors are equal to each other and have the form 
\begin{eqnarray}
 F_2(s) &=& -\biggl(1+ \frac{s-4 m_t^2}{12 m_t^2} \biggr)^{-1}.
 \label{eq:28}
\end{eqnarray}
This dependence is shown in Fig.~ \ref{fig:5}. 
As can be seen from Eq.~(\ref{eq:28}), the absolute value of the form factor is equal to 1 
at the reaction threshold $s=4 m_t^2$, and with increasing invariant energy the form factor decreases.

Finally, we consider experimental feasibility to observe the BSM contribution to the top-quark polarization (Table \ref{tab:2}).  
Assuming that the deviation of the averaged polarization component from its value in the SM is bigger than 
3 standard deviations,  
\begin{equation}
|\langle P_i \rangle - \langle P_i \rangle^{SM}| > 3 \frac{\sqrt{1 -\langle P_i \rangle^2}}{\sqrt{\mathcal{N}}},
\label{eq:29}
\end{equation}  
one can find the number of events $\mathcal{N}$ and the corresponding integrated luminosity $L = \int {\cal L} \, {\rm d} t$  \
needed to observe the BSM contribution at a 3$\sigma$ level with ideal detector. 
For a conservative estimate we choose values of $\kappa$ and $\kappa_z$ which are 10 times bigger than 
the radiative corrections (\ref{eq:16}).  Since the sign of the couplings is not certain, the  
negative signs of couplings are also considered. In addition, we present results for the values of $\kappa$ 
and $\kappa_z$ corresponding to the minimum of the unpolarized cross section (see the last line in Table \ref{tab:2}) .  
Note that a very small component $\langle P_y \rangle $ is not shown.

\begin{table}[tph]
\tbl{Anomalous couplings, transverse and longitudinal components $\langle P_x \rangle $ and $\langle P_z \rangle $, 
approximate numbers of events  $\mathcal{N}_x$ and $\mathcal{N}_z$, integrated luminosities   $L_x$ and $L_z$. 
The energy is $\sqrt{s}= 380$ GeV.  In the 1st line the SM values of polarization are shown.}
{\begin{tabular}{@{}c c c c c  c c c @{}}\toprule
& & \multicolumn{2}{c}{$\langle P \rangle$} & \multicolumn{2}{c}{$\mathcal{N}$} & \multicolumn{2}{c}{$L$, fb$^{-1}$} 
\\ \cmidrule(lr){3-4} \cmidrule(lr){5-6} \cmidrule(lr){7-8}
$\kappa$ & $\kappa_z$ & $\langle P_x \rangle$  & $\langle P_z \rangle$ & $\mathcal{N}_x$ & $\mathcal{N}_z$ & $L_x$ & $L_z$ \\ 
\midrule
0.02 & 0.00575 & -0.298 (SM) & -0.083 (SM) & -- & -- & -- & -- \\
0.2 & 0.0575 &  -0.307 & -0.065 & $  1.1 \times 10^5 $  & $  2.7 \times 10^4 $ & $123$ & $31$ \\
-0.2 & -0.0575 &  -0.274  & -0.123 & $  1.4 \times 10^4 $   & $ 5.3 \times 10^3 $   &  $56$ & $21$\\ 
-0.624 & -0.179 &  0.059  & -0.135 & $  0.7 \times 10^2 $   & $ 3.3 \times 10^3 $   &  $3$ & $151$\\ 
\botrule
\end{tabular} \label{tab:2}}
\end{table}


\section{Conclusions \label{sec:conclusions}}

The process of electron-positron annihilation into a pair of top quarks is considered in the conditions
 of the future electron-positron colliders. We chose the invariant energy $\sqrt{s}=380$ GeV which corresponds to 
 the planned conditions of the first run of CLIC. The components of the top-quark polarization are 
calculated in the SM and BSM in the framework of the
effective field theory in which additional $\gamma t \bar{t}$ and $Z t \bar{t}$ couplings are generated 
from the dimension-6 operators.
For the values of couplings $\kappa$ and $\kappa_z$, describing new physics,
we use the existing experimental bounds from Refs.~\citen{Bouzas:2013jha, Bouzas:2015} which are obtained from analysis 
of data on $B$-meson decays $B \to X_s \gamma$ and experiments at the LHC and the Tevatron 
on the top-quark production and decays. 
 
The dependence of the quark polarization components on the scattering angle is calculated 
for various values of $\kappa$ and $\kappa_z$. 
Our calculation shows that the sensitivity of the single-spin observables to the BSM couplings is sizable. 

The cross section for unpolarized quarks and the components of polarization averaged over scattering angle 
 are studied as functions of $\kappa$ and $\kappa_z$. 
The extreme points (maxima, minima and saddle points) of the corresponding two-dimensional surfaces  
are found and analyzed. Under an assumption that the true values of the couplings $\kappa$ and $\kappa_z$ may correspond to
position of the minimum of the cross section for unpolarized top quarks, we obtained the energy dependence of the form factors 
in the $\gamma t \bar{t}$ and $Z t \bar{t}$ vertices.  
 
In real experiments the top-quark polarization can be determined from measurements of angular distributions 
of the decay products of the top quark ($t \to W^+ \, b \to \ell^+ \, \nu_\ell \, b$). 
We are planning to address this aspect in future and also to extend calculations to the case of polarized 
electron and positron beams.


\section*{Acknowledgments}

Authors acknowledge partial support by the Ministry of Education and Science of Ukraine (Project No. 0117U004866) 
and the National Academy of Sciences of Ukraine (Project KPKVK  6541230). This research was partially conducted 
in the scope of the IDEATE International Associated Laboratory (LIA).

\appendix

\section{Matrix Element of $e^+ e^- \to t \, \bar{t}$,  Cross Section and Polarization Components \label{app:A}} 

The matrix element can be written as a sum of the SM and BSM terms
\begin{eqnarray}
M &=& M^{SM} + M^{BSM}.
\label{eq:A1}
\end{eqnarray}

For the  photon exchange the matrix elements are 
\begin{eqnarray}
iM_\gamma^{SM} &=& -i e^2 Q_t \frac{g_{\mu \nu}}{q^2}  \,
 \bar{u}(p) \gamma^{\mu} v(p^\prime) \, \bar{v}(k^\prime) \gamma^{\nu} u(k),
\nonumber \\
iM_\gamma^{BSM} &=& e^2 \frac{g_{\mu \nu}}{q^2} \,
 \bar{u}(p) \bigl( \kappa \, \frac{\sigma^{\mu \lambda}}{2 m_t}q_\lambda  \bigr) v(p^\prime)  \,
\bar{v}(k^\prime) \gamma^{\nu} u(k),
\label{eq:A3}
\end{eqnarray}
where $q=k+k^\prime=p+p^\prime$, \ $\sigma^{\mu \nu}=\frac{i}{2} \bigl(\gamma^\mu \gamma^\nu -\gamma^\nu \gamma^\mu \bigr)$, \ 
$g_{\mu \nu}$ is metric tensor with the signature $(+,-,-,-)$, \ $u(k)$, $v(k^\prime)$ are the Dirac spinors for electron and positron, and 
$u(p)$, $v(p^\prime)$ are the spinors for the quark and antiquark.  The invariant energy squared is $q^2 \equiv s = 4 E^2$.

For the Z-boson exchange the matrix elements are 
\begin{eqnarray}
iM_z^{SM} &=& i \frac{g^2}{4 c^2_w}\, D_{\mu \nu} (q) \,
 \bar{u}(p) \gamma^{\mu} \bigl(v_t -a_t \gamma^5 \bigr) v(p^\prime) \, \bar{v}(k^\prime) \gamma^{\nu} \bigl(v_e -a_e \gamma^5 \bigr) u(k),
\nonumber \\
iM_z^{BSM} &=& -\frac{g^2}{4 c^2_w} \, D_{\mu \nu} (q) \,
 \bar{u}(p)\bigl(\kappa_z \, \frac{\sigma^{\mu \lambda}}{2 m_t}q_\lambda  \bigr) v(p^\prime) \, \bar{v}(k^\prime) \gamma^{\nu} 
\bigl(v_e-a_e \gamma^5 \bigr) u(k).
\label{eq:A5}
\end{eqnarray}
In Eqs.~(\ref{eq:A5}) $D_{\mu \nu}(q) =({g_{\mu \nu}-q_\mu q_\nu/m^2_z }) /({q^2-m^2_z+i m_z\Gamma_z})$ 
and $\Gamma_z$ is the width of the Z boson.

Using the polarization density matrices  (\ref{eq:18}) we find the cross section for the unpolarized quarks
\begin{eqnarray}
\frac{d \sigma_0}{d \Omega} (\theta) &=& \frac{3 \alpha^2_{em} (s) V}{64 m_t^2 E^4 c^4_w s^4_w \Bigl(\Gamma_z^2 m_z^2+\bigl(m_z^2 - 4 E^2 \bigr)^2 \Bigr)}
\biggl[2 c^4_w s^4_w \Bigl(\Gamma_z^2 m_z^2 + \bigl(m_z^2 - 4 E^2 \bigr)^2 \Bigr) \nonumber \\
&& \times \Bigl(m_t^4 Q_t^2 + m_t^2 E^2 (3 \kappa^2 + 3 Q_t^2 + 8 \kappa Q_t) + V^2 E^2 \cos{2 \theta} (m_t Q_t - \kappa E) \nonumber \\
&& \times (m_t Q_t + \kappa E) + \kappa^2 E^4 \Bigr) - 4 E^2 c^2_w s^2_w (4 E^2 - m_z^2) \Bigl(4 m_t^2 V E^2 a_e a_t \cos{\theta} (\kappa + Q_t) \nonumber \\
&& + v_e (m_t^2 v_t (m_t^2 Q_t + Q_t V^2 E^2 \cos{2 \theta} + E^2 (4 \kappa + 3 Q_t)) + \kappa_z E^2 (m_t^2 (3 \kappa + 4 Q_t) \nonumber \\
&& - \kappa V^2 E^2 \cos{2 \theta} + \kappa E^2))\Bigr) + 2 E^6 \Bigl(V^2 \cos{2 \theta} (a_e^2 + v_e^2) (m_t^2 (a_t^2 + v_t^2) -  \kappa_z^2 E^2)\nonumber \\
&& + (a_e^2 + v_e^2) (m_t^2 (3 V^2 a_t^2 + v_t (8 \kappa_z - (V^2 - 4) v_t)) + \kappa_z^2 (3 m_t^2 + E^2)) \nonumber \\
&& + 16 m_t^2 V a_e a_t v_e \cos{\theta} (v_t + \kappa_z)\Bigr)\biggr],
\label{eq:A6}
\end{eqnarray}
and components of the polarization vector
\begin{eqnarray}
P_x &=& \frac{1}{N} \frac{E \sin{\theta}}{4 m_t c^4_w s^4_w \Bigl(\Gamma_z^2 m_z^2 + \bigl(m_z^2 - 4 E^2 \bigr)^2 \Bigr)}
\biggl[2 c^2_w s^2_w \bigl( 4 E^2 - m_z^2 \bigr) \Bigl( V a_t v_e \cos{\theta}  \nonumber \\
&& \times \bigl(m_t^2 Q_t + \kappa E^2 \bigr) + a_e \bigl(v_t \left(m_t^2 (\kappa + 2 Q_t) + \kappa E^2 \right) + \kappa_z \left(m_t^2 Q_t + E^2 
(2 \kappa + Q_t)\right)\bigr)\Bigr) \nonumber \\ 
&& - 2 E^2 \bigl( m_t^2 v_t + \kappa_z E^2 \bigr) \Bigl(2 a_e v_e \bigl( v_t + \kappa_z \bigr) + V a_t \cos{\theta} \bigl(a_e^2 + v_e^2 \bigr)\Bigr)\biggr],
\nonumber \\  
P_y &=& - \frac{1}{N} \frac{V E \Gamma_z m_z \sin{\theta} \Bigl(a_e a_t \bigl( m_t^2 Q_t + \kappa E^2 \bigr) + V E^2 v_e 
\cos{\theta} \bigl( \kappa v_t - Q_t \kappa_z \bigr)\Bigr)}{2 m_t c^2_w s^2_w \Bigl(\Gamma_z^2 m_z^2 + \bigl(m_z^2 - 4 E^2 \bigr)^2 \Bigr)},
\nonumber 
\\ 
P_z &=& - \frac{1}{N} \frac{E^2}{4 c^4_w s^4_w \Bigl(\Gamma_z^2 m_z^2 + \bigl(m_z^2 - 4 E^2 \bigr)^2 \Bigr)}
\biggl[E^2 \Bigl(4 a_e v_e \cos{\theta} \bigl( V^2 a_t^2 + \left(v_t + \kappa_z \right)^2 \bigr) \nonumber \\
&& + V a_t \bigl(\cos{2 \theta} + 3 \bigr) \bigl(a_e^2 + v_e^2 \bigr) \bigl(v_t + \kappa_z \bigr)\Bigr) - \bigl( \kappa + Q_t \bigr) c^2_w s^2_w 
\bigl(4 E^2 - m_z^2\bigr) \nonumber \\
&& \times \Bigl(4 a_e \cos{\theta} \bigl( v_t + \kappa_z\bigr ) + V a_t v_e \bigl(\cos{2 \theta} + 3\bigr)\Bigr)\biggr] 
\label{eq:A8} 
\end{eqnarray}
with $N$ defined in Eq.~(\ref{eq:23}).

\section{Averaging the Polarization Components over the Scattering Angle \label{app:B}}
In this Appendix we present the expressions for the averaged components of the top-quark polarization 
in Eq.~(\ref{eq:25}). These components are shown in Fig.~\ref{fig:4}. 
The obtained expressions are functions of the coupling constants $\kappa$ and $\kappa_z$.
The components read:
\begin{eqnarray}
\langle P_x \rangle &=& -\frac{1}{\langle N \rangle} \frac{\pi E a_e}{8 m_t c^4_w s^4_w 
\Bigl(\Gamma_z^2 m_z^2+\bigl(m_z^2 - 4 E^2 \bigr)^2 \Bigr)} \biggl[c^2_w s^2_w \bigl(m_z^2 - 4 E^2 \bigr) \nonumber \\ 
&& \times \Bigl(v_t \bigl(m_t^2 (\kappa + 2 Q_t) +  \kappa E^2 \bigr) + \kappa_z \bigl(m_t^2 Q_t + E^2 (2 \kappa + Q_t)\bigr)\Bigr) \nonumber \\ 
&& + 2 E^2 v_e \bigl(v_t + \kappa_z \bigr) \Bigl(m_t^2 v_t + \kappa_z E^2 \Bigr)\biggr],
\label{eq:B1}
\\
\langle P_y \rangle &=& -\frac{1}{\langle N \rangle} \frac{\pi V E \Gamma_z a_e a_t m_z 
\Bigl(m_t^2 Q_t+\kappa E^2\Bigr)}{8 m_t c^2_w s^2_w \Bigl(\Gamma_z^2 m_z^2+\bigl(m_z^2-4 E^2 \bigr)^2 \Bigr)},
\label{eq:B2} 
\\
\langle P_z \rangle &=& -\frac{1}{\langle N \rangle} \frac{2 V E^2 a_t}{3 c^4_w s^4_w \Bigl(\Gamma_z^2 m_z^2+\bigl(m_z^2-4 E^2 \bigr)^2 \Bigr)} 
\biggl[E^2 a_e^2 \bigl(v_t + \kappa_z \bigr) \nonumber \\ 
&& + v_e \Bigl(c^2_w s^2_w \bigl(m_z^2-4 E^2 \bigr) \bigl(\kappa + Q_t \bigr)  + E^2 v_e \bigl(v_t + \kappa_z \bigr)\Bigr)\biggr].
\label{eq:B3}
\end{eqnarray}
The dimensionless constant $\langle N \rangle $ is related to the total cross section for unpolarized quarks, 
\begin{equation}  
\langle N \rangle  =  \frac{E^2 \, \sigma_0}{3 \, \alpha_{em}^2 (s) \,  \pi \,  V}, 
\label{eq:B4}
\end{equation}  
\begin{eqnarray}
\sigma_0 &=& \frac{\pi  \alpha^2_{em} (s) V}{2 m_t^2 E^4 c^4_w s^4_w \Bigl(\Gamma_z^2 m_z^2+\bigl(m_z^2 - 4 E^2 \bigr)^2 \Bigr)}
\biggl[c^4_w s^4_w \Bigl(\Gamma_z^2 m_z^2 + \bigl(m_z^2 - 4 E^2 \bigr)^2 \Bigl) \nonumber \\
&& \times \Bigl(m_t^4 Q_t^2 + 2 m_t^2 E^2 \left(\kappa^2 + Q_t^2 + 3 \kappa Q_t\right)+ \kappa^2 E^4\Bigr) + E^6 \bigl(a_e^2 + v_e^2 \bigr) \nonumber \\
&& \times \Bigl(m_t^2 \bigl(2 V^2 a_t^2 + v_t \left((3 - V^2) v_t + 6 \kappa_z \right)\bigr) + 
\kappa_z^2 \left(2 m_t^2 + E^2 \right)\Bigr) - 2 E^4 v_e c^2_w s^2_w  \nonumber \\
&& \times \bigl(4 E^2 - m_z^2 \bigr) \Bigl(m_t^2 v_t \bigl(3 \kappa + Q_t (3 - V^2)\bigr) + \kappa_z \bigl(m_t^2 (2 \kappa + 3 Q_t) + \kappa E^2\bigr)\Bigr)\biggr].
 \label{eq:B5}
\end{eqnarray}

As can be seen from these equations, the components of the polarization are complicated functions of the 
coupling constants $\kappa$ and $\kappa_z$. 



\begin{thebibliography}{0}    

\bibitem{Tanabashi:2018oca}
M.~Tanabashi {\it et al.} [ParticleDataGroup],
Phys.\ Rev.\ D {\bf 98}, 030001 (2018).

\bibitem{Chatrchyan:2013wua} 
  S.~Chatrchyan {\it et al.} [CMS Collaboration],
  Phys.\ Rev.\ Lett.\  {\bf 112}, 182001 (2014).

\bibitem{Aad:2013ksa} 
  G.~Aad {\it et al.} [ATLAS Collaboration],
  Phys.\ Rev.\ Lett.\  {\bf 111}, 232002 (2013).

\bibitem{Khachatryan:2016xws} 
  V.~Khachatryan {\it et al.} [CMS Collaboration],
  Phys.\ Rev.\ D {\bf 93}, 052007 (2016).

\bibitem{deBlas:2018mhx} 
  J.~de Blas {\it et al.},
  CERN-TH-2018-267, CERN-2018-009-M, 
  arXiv:1812.02093 [hep-ph].
	
	\bibitem{Roloff:2018dqu} 
  P.~Roloff {\it et al.} [CLIC and CLICdp Collaborations],
  arXiv:1812.07986 [hep-ex].
	
\bibitem{Vos:2016til} 
  M.~Vos {\it et al.}, 
  arXiv:1604.08122 [hep-ex].

\bibitem{Garcia:2016}
I. Garcia Garcia,
CERN-THESIS-2016-214. 
	
\bibitem{CLIC:2016zwp}
  M.~J.~Boland {\it et al.} [CLIC and CLICdp Collaborations],
  arXiv:1608.07537 [physics.acc-ph].

\bibitem{Buckley:2017} 
A.~Buckley, C.~Englert, J.~Ferrando, D.~J.~Miller, L.~Moore, M.~Russell and C.~D.~White,
JHEP {\bf 1604}, 015 (2016); and update in arXiv:1512.03360 [hep-ph].

\bibitem{Bouzas:2013jha} 
  A.~O.~Bouzas and F.~Larios,
  Phys.\ Rev.\ D {\bf 88}, 094007 (2013).
	
\bibitem{Bouzas:2015}
A.~O.~Bouzas and F.~Larios,
J.\ Phys.\ Conf.\ Ser.\  {\bf 651}, 012004 (2015).

\bibitem{Burges:1983} 
C.~J.~C.~Burges and H.~J.~Schnitzer,
Nucl.\ Phys.\ B {\bf 228}, 464 (1983).

\bibitem{Leung:1984} 
C.~N.~Leung, S.~T.~Love and S.~Rao,
Z.\ Phys.\ C {\bf 31}, 433 (1986).

\bibitem{Buchmuller:1985} 
W.~Buchmuller and D.~Wyler,
Nucl.\ Phys.\ B {\bf 268}, 621 (1986).

\bibitem{AguilarSaavedra:2008} 
J.~A.~Aguilar-Saavedra,
Nucl.\ Phys.\ B {\bf 812}, 181 (2009),  

\bibitem{Grzadkowski:2010} 
B.~Grzadkowski, M.~Iskrzynski, M.~Misiak and J.~Rosiek,
JHEP {\bf 1010}, 085 (2010).  

\bibitem{Hollik:1998vz} 
  W.~Hollik, J.~I.~Illana, S.~Rigolin, C.~Schappacher and D.~Stockinger,
  Nucl.\ Phys.\ B {\bf 551}, 3 (1999);   Erratum: Nucl.\ Phys.\ B {\bf 557}, 407 (1999).

\bibitem{Czarnecki:1995sz} 
A.~Czarnecki, B.~Krause and W.~J.~Marciano,
Phys.\ Rev.\ Lett.\  {\bf 76}, 3267 (1996). 

\bibitem{Hollik:1988ii} 
W.~F.~L.~Hollik,
Fortsch.\ Phys.\  {\bf 38}, 165 (1990).

\bibitem{Bernabeu:1997je} 
J.~Bernabeu, J.~Vidal and G.~A.~Gonzalez-Sprinberg,
Phys.\ Lett.\ B {\bf 397}, 255 (1997). 

\bibitem{Bernreuther:2004ih} 
W.~Bernreuther, R.~Bonciani, T.~Gehrmann, R.~Heinesch, T.~Leineweber, P.~Mastrolia and E.~Remiddi,
Nucl.\ Phys.\ B {\bf 706}, 245 (2005). 

\bibitem{Bernreuther:2004th} 
W.~Bernreuther, R.~Bonciani, T.~Gehrmann, R.~Heinesch, T.~Leineweber, P.~Mastrolia and E.~Remiddi,
Nucl.\ Phys.\ B {\bf 712}, 229 (2005). 
	
\bibitem{Bernreuther:2005rw} 
W.~Bernreuther, R.~Bonciani, T.~Gehrmann, R.~Heinesch, T.~Leineweber and E.~Remiddi,
Nucl.\ Phys.\ B {\bf 723}, 91 (2005). 
	
\bibitem{Bernreuther:2005gq} 
W.~Bernreuther, R.~Bonciani, T.~Gehrmann, R.~Heinesch, T.~Leineweber, P.~Mastrolia and E.~Remiddi,
Phys.\ Rev.\ Lett.\  {\bf 95}, 261802 (2005). 

\bibitem{Arens:1992}
T.~Arens and L.~M.~Sehgal,
Nucl.\ Phys.\ B {\bf 393}, 46 (1993).

\end{thebibliography}
\end{document}